\documentclass[conference]{IEEEtran}

\usepackage{cite}
\usepackage{amsmath}
\usepackage{amsmath,amssymb,amsfonts}
\usepackage{algorithmic}
\usepackage{graphicx}
\usepackage{textcomp}
\usepackage{xcolor}
\usepackage{pdfpages}
\usepackage{textcomp}
\usepackage{tikz}
\usepackage{xfrac}
\usepackage{xcolor}
\usepackage{amssymb}
\usepackage{mathtools}
\usepackage{amsmath}
\usepackage{enumerate}
\usepackage{blindtext}
\usepackage{enumitem}
\usepackage{graphicx}
\usepackage{caption}
\usepackage{subcaption}
\graphicspath{{Figures/}}
\usepackage[justification=centering]{caption}
\usepackage{setspace}
\usepackage{bbm}
\usepackage{listings}
\usepackage{multirow}

\usepackage{booktabs}

\newcommand{\bftab}{\fontseries{b}\selectfont}

\begin{document}
	
	\title{ML-Aided Collision Recovery for UHF-RFID Systems \vspace{-4mm}}	
	\author{\IEEEauthorblockN{Talha Aky{\i}ld{\i}z\IEEEauthorrefmark{1}, Raymond Ku\IEEEauthorrefmark{1}, Nicholas Harder\IEEEauthorrefmark{1}, Najme Ebrahimi\IEEEauthorrefmark{2}, Hessam Mahdavifar\IEEEauthorrefmark{1}} 	
		\IEEEauthorblockA{\IEEEauthorrefmark{1} Department of Electrical Engineering and Computer Science, University of Michigan, Ann Arbor, MI 48109, USA}
		\IEEEauthorblockA{\IEEEauthorrefmark{2} Department of Electrical and Computer
			Engineering, University of Florida, Gainesville, FL 32603, USA} 
		Email: \{akyildiz, rayku, nharder\}@umich.edu,	najme@ece.ufl.edu, hessam@umich.edu
		\vspace{-4mm}	
		 
	}
	\maketitle
	\normalsize

	\begin{abstract}	
		We propose a collision recovery algorithm with the aid of machine learning (ML-aided) for passive Ultra High Frequency (UHF) Radio Frequency Identification (RFID) systems. The proposed method aims at recovering the tags under collision to improve the system performance. We first estimate the number of tags from the collided signal by utilizing machine learning tools and show that the number of colliding tags can be estimated with high accuracy. Second, we employ a simple yet effective deep learning model to find the experienced channel coefficients. The proposed method allows the reader to separate each tag's signal from the received one by applying maximum likelihood decoding. We perform simulations to illustrate that the use of deep learning is highly beneficial and demonstrate that the proposed approach boosts the throughput performance of the standard framed slotted ALOHA (FSA) protocol from $0.368$ to $1.837$, where the receiver is equipped with a single antenna and capable of decoding up to $4$ tags. 		
	\end{abstract}
	
	\section{Introduction}
	
	Passive Ultra High Frequency (UHF) Radio Frequency Identification (RFID) is a wireless communication system. UHF-RFID systems have a wide range of application area, e.g., logistics \& supply chain, item inventory tracking and materials management to mention a few.  In UHF-RFID systems, the communication is conducted between the reader and arbitrary number of passive tags with backscatter modulation. In this work, we focus on passive UHF-RFID systems where the tags can only operate by absorbing energy from the reader through the radio waves and have a limited operation capability.
	
	In UHF-RFID systems, the reader can only communicate with at most one tag, and whenever two or more tags try to communicate with the reader a collision occurs. To avoid that, EPC Gen2 standards employ framed slotted ALOHA (FSA) protocol to randomize the transmission procedure of passive tags \cite{global2008epc}. In FSA, frames are formed by slots and each tag chooses its slot uniformly over the frame where its size is determined by the reader and also known by the tags \cite{schoute1983dynamic}. Even though FSA is employed as an anti-collision protocol, collisions still occur. This situation leads to the application of collision recovery algorithms with an aim of extracting or possibly recovering information from the tags under collision.
	
	In \cite{shen2009separation}, the authors study joint-decoding of multiple tags by investigating the signal constellation points and parameter estimation methods for low frequency RFID systems. Another work \cite{khasgiwale2009extracting} shows that the number of tags up to four can be estimated from the collided signal, and this information later can be used to recover the signals of colliding tags. The utilization of fixed beam-forming for different sub-spaces to detect collisions which can be combined with other anti-collision algorithms is presented in \cite{yu2008anti}. In \cite{mindikoglu2008separation}, it is shown that using antenna array which is coupled with blind source separation techniques can reduce the collision probability and also remove the interference signals. Another related work \cite{knerr2010slot} formulates a maximum likelihood estimator to find out the number of tags in collision for each slot and also compare its performance against to the other well known estimators.
	
	The authors in \cite{angerer2010single} first propose a different type of single antenna collision recovery receivers, i.e., zero-forcing and ordered successive cancellation, to improve the performance of FSA by estimating channel coefficients for at most two colliding tags, and then, they generalize their approach for multiple antenna setting in \cite{angerer2010rfid}. Another similar work \cite{kaitovic2011rfid} shows that it is possible to recover more than two colliding tags by exploiting the additional diversity (increasing the number of receiver antennas) with an assumption of perfect channel knowledge at the reader. Later, they extend their work by also proposing a channel estimation technique using the post-preamble symbols in \cite{kaitovic2012channel}. In \cite{tan2016collision}, the number of tag estimation is coupled with a channel estimation technique for collision recovery and \cite{mahdavifar2015coding} studies code designs to recover tag collisions given the number of colliding tags.
	
	In this work, we propose a new framework with the application of deep learning tools for the collision recovery in UHF-RFID systems. In our framework, we first estimate the number of tags in collision by considering Gaussian mixture model (GMM) which clusters the received signal points. We then replace GMM with two different neural networks architectures, i.e., feed-forward and convolutional. We observe that the deep learning architectures can estimate the number of tags with high accuracy by out-performing GMM. The estimated number of tags is then utilized to identify the channel coefficients of tags with the aid of additional symbols via deep learning. As a last step, we decode the backscattered modulated signals using minimum-distance decoding. The results demonstrate that the proposed framework achieves significantly higher throughput values compared to the conventional schemes, and hence, the use of deep learning tools is highly beneficial for collision recovery in UHF-RFID systems. 
	
	The remainder of this paper is organized as follows. In Section II, we present the system model including FSA and the communication channel. Section III describes the machine learning based proposed model for collision recovery. Section IV illustrates our numerical results in comparison with the other approaches. Finally, Section V concludes the paper.

	\section{System Model}
	In this section, we describe a mathematical framework for UHF-RFID systems. First, we briefly mention the properties of FSA and also compute the analytical throughput performance of FSA with collision recovery. We also represent the communication model between the reader and passive tags. Finally, we depict the constellation points of the received signal at the reader in in-phase (I) and quadrature (Q) plane (I/Q plane). 
	
	\subsection{Framed Slotted ALOHA}
	EPC Gen2 standards employ FSA protocol where frames are formed by $K$ slots and a population of $N$ tags select their slots uniformly over the frame. In FSA, slots can be in three different states: 1) Empty slot: None of the tags transmit, 2) Singleton Slot: Only one specific tag transmits, 3) Collision Slot: Two or more tags try to communicate. In conventional UHF-RFID systems, the reader can only identify singleton slots and collision slots are discarded. Let $\mathcal{X}_R$ be a random variable that denote the number of slots with $R$ colliding tags and its expected value is $\mathbb{E}\{\mathcal{X}_R\} = K \binom{N}{R} \left(\frac{1}{K}\right)^{R} \left(1-\frac{1}{K}\right)^{N-R}$.
		
	
	The throughput of FSA is defined as the number of decoded users per slot which corresponds to $\frac{\mathbb{E}\{\mathcal{X}_1\}}{K}$. This value is maximized when $K=N$ with a value of $0.368$.
	
	We consider the scenario where the receiver is capable of decoding tags when collision occurs and derive the throughput under this scenario. Let $J$ denote the number of tags can be decoded and $M$ denote the number of tags can be resolved by the receiver with a condition $R \leq M$. If $J=1$, the receiver can decode one of the tags out of at most $M$ colliding ones. In this case, the throughput can be computed as $\frac{1}{K} \sum_{R=1 }^{M} \mathbb{E} \{\mathcal{X}_R\} =   \sum_{R=1 }^{M} \binom{N}{R} \left(\frac{1}{K}\right)^{R} \left(1-\frac{1}{K}\right)^{N-R}$.
	
\begin{figure}[h]
		\centering
		\includegraphics[clip, trim=0.7cm 0.12cm 1.8cm 1.1cm,width=1\columnwidth]{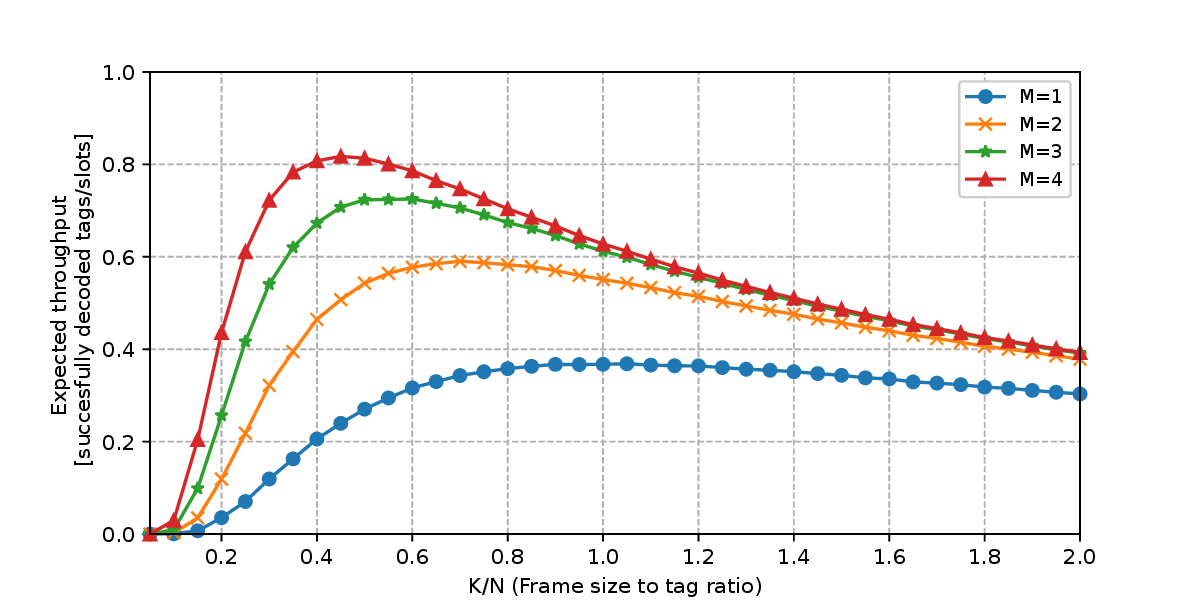}
		\caption{Expected throughput comparison for different values of $M$ over different values of $K/N$ when $J=1$.}
		\label{fig1}
		\vspace{-0.1cm}
	\end{figure}

	We illustrate the throughput values of FSA over different values of frame size to tag ratio ($K/N$) for  $M$ up to $4$ in Fig. \ref{fig1}. It can be seen that as $M$ increases, the throughput performance improves, and the maximum throughput can be achieved for $K/N$ values because the collisions are now recoverable. The throughput can be increased to $0.818$ ($M=4$) which is a substantial gain compared to the throughput $0.368$ of FSA.		 
	
	We also consider the scenario where the receiver can decode more than one tag ($J \geq 1$) out of up to $M$ colliding tags. In this case, the throughput value can be written as $\frac{1}{K} (\sum_{R=1 }^{J} R ~ \mathbb{E} \{\mathcal{X}_R\} + \sum_{R=J+1 }^{M} J ~ \mathbb{E} \{\mathcal{X}_R\} )$.
	
\begin{figure}[h]
		\centering
		\includegraphics[clip, trim=0.7cm 0.12cm 1.8cm 1.1cm,width=1\columnwidth]{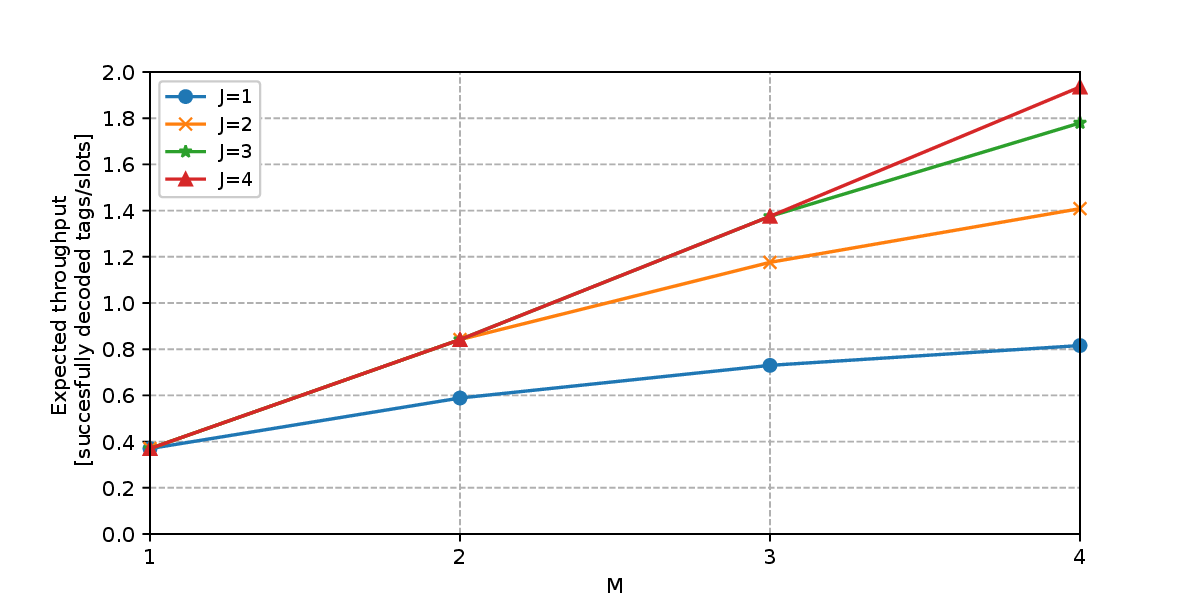}
		\caption{Expected throughput comparison for different values of $M$ and $J$ with optimal frame size to tag ratio ($K/N$).}
		\label{fig2}
		\vspace{-0.4cm}
	\end{figure}

	We depict the maximum throughput values of FSA for different values of $M$ and $J$ in Fig. \ref{fig2} by calculating the optimal $K/N$ values. It can be seen that recovering more than one tag is significantly beneficial and can boost the system performance further. The throughput value can be up to nearly $2$ which was $0.818$ when $J=1$. The throughput analysis of FSA with collision recovery shows its potential and how it can improve the performance of conventional FSA. \vspace{-0.1cm}

	\subsection{The Communication Model}
\begin{figure}[t]
		\centering
		\includegraphics[clip, trim=3cm 6cm 8.5cm 3.5cm,width=0.9\columnwidth]{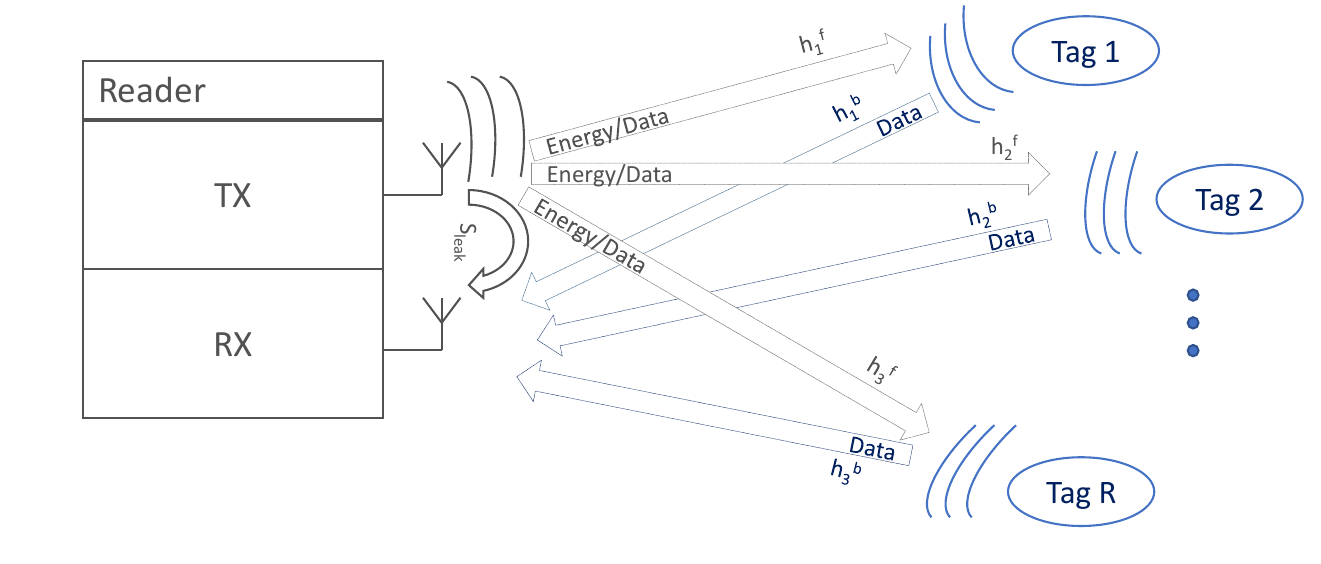}
		\caption{The communication model between the reader and passive tags of number $R$.}
		\label{fig3}
		\vspace{-0.68cm}
	\end{figure}

	We now present the communication model between the reader and passive tags, following typically considered models in the system level analysis of UHF-RFID systems \cite{angerer2010rfid,bletsas2012single} as the models also comply with measurement data. The communication model is depicted in Fig. \ref{fig3}. In our model, we only consider a single receiver antenna at the reader. The communication can be divided into two sub-parts: 1) Forward channel: The reader supports energy and data to the passive tags with continuous carrier wave transmission, 2) Backward channel: Passive tags absorb the energy from the reader and reflects it back to the reader. Each tag modulates its 16 bit random number (RN16) with an on-off keying: 0 (OFF) and 1 (ON) corresponding to the absorbing and reflecting states. Let $a_i(t)$ denote the modulated RN16 signal for tag $i$, i.e.,
	\begin{equation}
		a_i(t) = \sum_{k} a_i[k] p(t-kT_i - \tau_i),
	\end{equation}
	where $a_i[k]$ denotes the transmitted symbols ($\pm 1$) and $p(t)$ is a rectangular pulse of the modulated signal. $T_i$ and $\tau_i$ correspond to the symbol period and modulation delay, respectively. 
	
	The backscattered modulated signal for tag $i$ is transmitted via carrier frequency $f_c$ and denoted by $s_{tag,i}$, i.e., 
	\begin{equation}
		s_{tag,i} = |h_i^f| \sqrt{|\Delta\sigma_i|} a_i(t) \sin(2\pi f_c t + \varphi_i^f + \varphi_i^{\Delta\sigma} ),
	\end{equation} 
	where $|h_i^f|$ is the magnitude of forward channel channel coefficient, $|\Delta\sigma_i|$ is the normalized differential radar cross section (RCS) coefficient of tag $i$ \cite{nikitin2007differential}. $\varphi_i^f$ and $\varphi_i^{\Delta\sigma}$ are phase shifts of the forward channel and the modulation.
	
	During the transmission from the reader to tags, a leakage occurs between the transmitter and receiver antenna, i.e.,
	\begin{equation}
		s_{leak}(t) = |L| \sin(2\pi f_c t + \varphi^{leak} ),
	\end{equation}
	where $|L|$ is leakage magnitude and $\varphi^{leak}$ is the phase shift.
	
	After deriving all necessary components, we  finally compute the received signal at the reader denoted as $s(t)$ and given by
	\begin{align*}
		s(t) &= \sum_{i=1}^{R} |h_i^b| |h_i^f| \sqrt{|\Delta\sigma_i|} a_i(t) \sin(2\pi f_c t + \varphi_i^f + \varphi_i^{\Delta\sigma} + \varphi_i^b )\\ &+ s_{leak}(t) + n(t),
	\end{align*}
	where $R$ is the number of colliding tags in a certain slot. The backward channel is denoted by $h_i^b$ and $\varphi_i^b$ is the corresponding phase shift for tag $i$. $n(t)$ is the additive white Gaussian noise (AWGN) with power spectral density $N_0$\footnote{We note that the parameters, i.e., the channel coefficients, phase shifts, differential RCS is constant during the transmission of the modulated backscattered tag signals.}. 
	
	It is possible to down-convert the received signal into the baseband using I/Q demodulators since all signal components have the same carrier frequency. Utilizing that, we can write the complex-valued baseband signal at the reader as
	\begin{equation}\label{eq1}
		s^b(t) =  \sum_{i=1}^{R} h_i^b h_i^f \sqrt{\Delta\sigma_i}a_i(t) + L + n^b(t),
	\end{equation}
	where  $h_i^f = |h_i^f|e^{\varphi_i^f}$ and $h_i^b = |h_i^b|e^{\varphi_i^b}$ are complex valued channel coefficients. The normalized RCS, leakage and noise can be written as $\sqrt{\Delta\sigma_i} = \sqrt{|\Delta\sigma_i|} e^{\varphi_i^{\Delta\sigma}}$,  $L = |L| e^{\varphi^{leak}}$, and $n^b(t) = n(t)e^{j 2\pi f_c t}$, respectively.
	
	We further simplify the equation (\ref{eq1}) by denoting $h_i = h_i^b h_i^f \sqrt{\Delta\sigma_i}$ as the general coefficient including both forward and backward channels as well as differential RCS coefficient. By introducing $\mathbf{h}$ with elements $h_i$, $\mathbf{a}(t)$ with elements $a_i(t)$, we can rewrite the received signal as
	\begin{equation}
		s^b(t) =   \mathbf{h}\mathbf{a}(t) + L + n^b(t),
	\end{equation}
	where signal to noise ratio (SNR) is defined as $\mathbb{E}\{ |h_i|^2 a_i^2  \}/ N_0$.
	
	We note that our communication model assumes coherent detection, i.e, precise tag synchronization and symbol rate among tags. In UHF-RFID systems, the tags might be asynchronous due to delay term ($\tau_i$) during the modulation and modulated RN16 signals $a_i(t)$ generally have different modulation frequency (symbol period $T_i$) which can differ up to $\pm 22\% $ \cite{global2008epc}. However, the impact of asynchronism and different symbol periods among tags is neglected to obtain comparable numerical results with the literature where simulations are performed in ideal conditions and non-coherent detection left as a future work.
	
	\subsection{Received Signal Constellation Points}
	
\begin{figure}
		\centering
		\includegraphics[clip, trim=0cm 0cm 0cm 0cm,width=1\columnwidth]{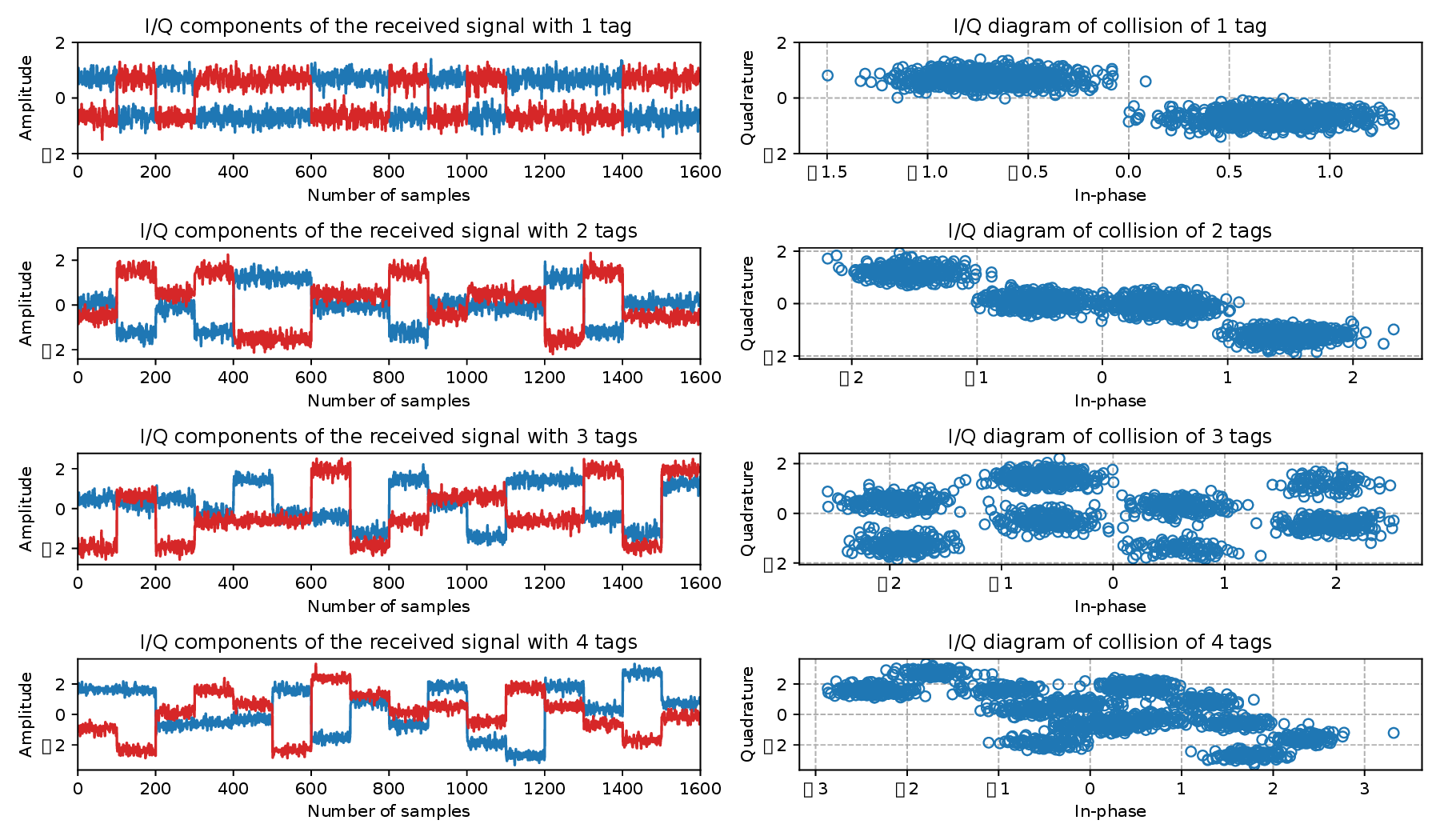}
		\caption{Illustration of the received signal levels and constellation points for different number of tags up to $4$ when SNR is $20$ dB.}
		\label{fig4}
		\vspace{-0.6cm}
	\end{figure}

	The received baseband signal is complex-valued and contains both in-phase  and quadrature components. The received signal $s^b(t)$ and its constellation points in I/Q plane for different values of number of tags up to $4$ is illustrated in Fig. \ref{fig4}. The first column plots correspond to the amplitudes of both I/Q components of the received signal where second column plots depict the constellation points in I/Q plane. One major observation is that for each number of tag we have different amplitude levels which results in higher number of clusters as we increase the number of tags. Since we only consider ON-OFF keying, the transmitted symbols can only take values $\pm 1$. As a result of that, the cluster centers and signal levels correspond to the distinct combinations of the channel gains, $h_i$, and therefore, we will have $2^R$ different levels and clusters when the number of tags in collision is $R$. 
	
	The I/Q diagram of the received signal provides important information about the number of tags under collision. Hence, we utilize the I/Q diagram of the constellation points to estimate the number of tags in the proposed recovery algorithm.
	
	\section{ML-Aided Collision Recovery Algorithm}
\begin{figure}[b]
		\centering
		\includegraphics[clip, trim=1cm 6cm 9cm 7cm,width=1\columnwidth]{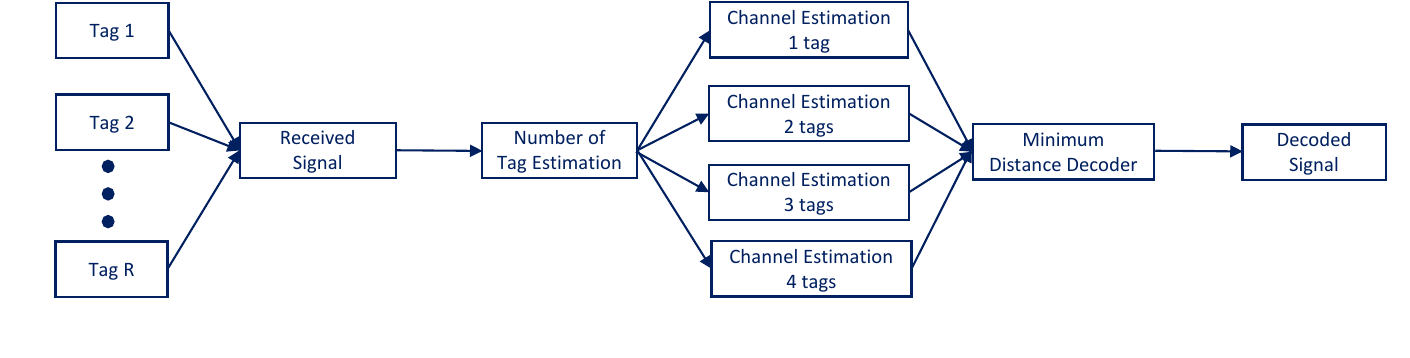}
		\caption{The workflow diagram of the proposed ML-aided collision recovery algorithm for number of tags up to $4$.}
		\label{fig5}
		\vspace{-0.6cm}
	\end{figure}

	In this section, we present our proposed ML-aided collision recovery algorithm for number of tags up to $4$ and the diagram of our algorithm is shown in Fig. \ref{fig5}. As a first step, we estimate the number of tags in collision by using GMM and neural network architectures, i.e., feed-forward (FNN) and convolutional (CNN), utilizing the received signal constellation points in I/Q plane. Then, four different FNN models are trained to estimate the channel coefficients for given number of tags with the aid of $4$ additional symbols. After finding the number of tags and the channel gains, we apply minimum distance decoder to separate the transmitted signals of the passive tags. 
	
	\subsection{Number of Tag Estimation}
	
	We first consider a well-known clustering method Gaussian mixture model (GMM) for number of tag estimation. GMM is a probabilistic model-based clustering technique where the samples are drawn from the mixture of normal distributions, i.e., $p(\mathbf{x_n}) = \sum_{l=1}^{L} {\pi_l} N(\mathbf{x_n} \mid \mu_l, \Sigma_l )$ where $\mathbf{x_n} $ is received signal constellation points, $\pi_l$ is mixture probabilities, i.e., $0 \leq \pi_l \leq 1$ for all $l$ and $\sum_{l=1}^{L} \pi_l = 1$,  $N(\mathbf{x_n} \mid \mu_l, \Sigma_l )$ is a normal distribution with a mean $\mu_l$ and a covariance $\Sigma_l$ where $L$ is the number of components in the mixture equal to $2^R$. GMM technique suits well for our problem since the received signal is a mixture of normally distributed constellation points and they are centered at the combinations of the channel coefficients and deviates around the center points with an AWGN noise\footnote{For GMM, we pick the number of tags by utilizing the estimated number of clusters. We select the number of tags as $x$ when the number of clusters lies between $2^{x-1}$ and $2^{x}$.}. We use expectation-maximization (EM) algorithm as an iterative solution to find a well-performing parameters ($\pi_l, \mu_l,{\Sigma_l}$) of GMM \cite{dempster1977maximum}. 
	
	We extend our approach by using FNN and the architecture model comprises of $7$ hidden layers and an output layer with the number of units starting from $1024$ and gradually decreasing through the hidden layers by half where the output layer has $4$ units corresponding to the number of tags to be estimated. The received baseband signal $s^b(t)$ is fed into the model by concatenating the real and imaginary parts. All linear layers apply a matrix multiplication with weight parameters and also add bias terms. After linear operations, a ReLU (rectified linear unit) activation function is applied at the hidden layers. At the output layer, soft-max activation function is used to get probabilities for the number of tags. 
	
	We also consider CNN with an aim of improving the performance of FNN since it might be able to extract features from the constellation points through the convolutional layers. We use similar architecture for CNN as in FNN. Different from FNN, we feed the received signal to the 3 convolutional layers with a number of channels $16$, $64$ and $32$ with a kernel size of $5$ and activated by ReLU function. The convolutional layers is then followed by $4$ hidden layers and an output layer with a number of units $512$, $256$, $64$, $32$, and $4$ respectively. Moreover, hidden and output layers are coupled with ReLU and soft-max activation functions.
	
	We describe the dataset used for the training of FNN and CNN. We generate the received signal samples for each number of tag in size of batches and symbol period $T_i$ and delay parameters $\tau_i$ are set to $10$ and $0$, respectively. The training SNR is determined as $5$ dB and leakage is assumed as 0. Each received signal sample is formed as follows: 1) Each tag modulates 16 bit random number (RN16), 2) The Rayleigh channel coefficients are generated as an independent and identically distributed (i.i.d) with zero mean and unit variance complex Gaussian random variables, i.e., $\mathcal{CN}(0,1)$, 3) The noise realizations are also i.i.d complex Gaussian random variables with zero mean and variance $N_0$. 
	
	For GMM, we apply Bayesian-based information criterion (BIC) score to avoid over-estimating the number of clusters by applying penalization over the likelihood function \cite{schwarz1978estimating}. For FNN and CNN, Adam optimizer \cite{adam2} is used with a learning rate $\eta = 0.0001$. We select the batch size as $256$ for each number of tag and employ cross-entropy loss function. 
	
	\subsection{Channel Estimation}
	We now explain our proposed approach for channel estimation by utilizing the estimated number of tags. After finding out the number of tags, four different FNN models are trained with the aid of $4$ fixed symbols in addition to the RN16. We use fixed additional symbols for this specific problem because if the symbols are formed in a random manner similar to the RN16, the neural network cannot capture the relationship between the received signal and the channel coefficients. We select the number of symbols as $4$ since we are interested in estimating the channel coefficients for at most $4$ tags. We note that the additional symbols should be different for each tag and orthogonal to each other. The main idea of using FNN model is to extract the channel coefficients from the received signal as it is corrupted by the AWGN noise. 
	
	The FNN model for channel estimation contains $5$ hidden layers with an output layer. Similar to the previous model, the received signal with only $4$ additional symbols is fed into the model. The hidden layers have equal number of $320$ units and are activated by ReLU function. The output layer has no activation function and consists of $2R$ units where multiplier $2$ is used for the real and imaginary parts. For the training procedure, we produce the dataset in a similar fashion to the number of tag estimation case. We use Adam optimizer with learning rate $\eta = 0.1$. We select the batch size as $256$ with a training SNR of $20$ dB and also employ minimum mean squared error (MMSE) as the loss function.
	
	\subsection{Minimum Distance Decoder}
	As a final step in our framework, we perform minimum distance decoding after finding the channel coefficients for each tag. As described earlier in Section II, the received signal levels or cluster centers are combinations of channel coefficients with different signs. As an example, consider two tags with channel gains $h_1$ and $h_2$, we will have $4$ different levels in the received signal denoted by $l_1$, $l_2$, $l_3$, and $l_4$, i.e.,
	\begin{equation*}
		l_1 = h_1 + h_2, ~~ l_2 = h_1 - h_2, ~~ l_3 = -h_1 + h_2,  ~~ l_4 = -h_1 - h_2.
	\end{equation*} 
	After computing the signal levels which can be done for higher number of tags in a similar manner, the maximum likelihood decision rule is equivalent to a minimum distance decoding for the given observation (received signal $s^b(t)$), i.e.,
	\begin{equation}
		\hat{i} = \underset{i \in \{1,2, \cdots, 2^R \}}{\mathrm{argmin}}{\mid\mid l_i - s^b(t) \mid\mid}
	\end{equation}
	As a final step, after finding the optimal signal level for each RN16 symbols, we can derive the transmitted bits/symbols of passive tags for the given signal level. Specifically, if $\hat{i} = 1$, both tags transmit $1$ in their reflecting states and if $\hat{i} = 2$, first and second tag transmit $1$ and $-1$, respectively.
	
	\section{Numerical Results}
	In this section, we illustrate the performance of the proposed framework via simulations. We first present the accuracy values of tag estimation and MMSE losses for channel estimation. We then demonstrate the throughput performance of GMM, FNN and CNN in comparison with the existing approaches. 
	
	We consider two frequency flat channel models, i.e., Rayleigh and Rician fading. In the former, the channel coefficients are i.i.d with zero mean and unit variance complex Gaussian random variables. In the latter, the channel coefficients are calculated as $h_{\text{ric}} = \sqrt{P_{LOS}} + \sqrt{P_{NLOS}} ~ h_{\text{ray}}$ where $h_{\text{ric}}$ and $h_{\text{ray}}$ are the corresponding channel coefficients. $P_{LOS}$ and $P_{NLOS}$ measure the powers of line of sight (LOS) and non line of sight (NLOS) components in the environment and can be computed as $P_{LOS} = \frac{K}{K+1}$ and $P_{NLOS} = \frac{1}{K+1}$ where $K$ is the Rician factor and selected as $2.8$ dB according to the measurements in \cite{kim2003measurements}.
	
	\subsection{Number of Tag and Channel Estimation}
	The accuracy values of GMM, FNN and CNN are given in Table \ref{tab2} for tag numbers up to $4$ at an SNR value of 20 dB under both channel models. While GMM performs well at SNR value of $20$ dB, its performance degrades severely for lower SNR values. On the other hand, FNN has the worst performance and CNN outperforms both GMM and FNN with an overall accuracy value of $0.971$ for Rayleigh fading channel. However, it should be noted that FNN and CNN are more tolerant to noise than GMM as it will be illustrated in throughput simulations. It is also possible to see that FNN and CNN can achieve almost identical performance for Rician fading model as well. This exhibits that even though the neural networks are trained with a specific channel model, it is still robust to other channel model as well.
	
	
	\begin{table}[h]
		\small
		\caption{Comparison of tag accuracy values for GMM, FNN and CNN at SNR value of 20 dB under both channel models.}
		\centering
		\renewcommand{\arraystretch}{1.2} 
		\resizebox{0.48\textwidth}{!}{
			\begin{tabular}{|c|c|c|c|c|c|}
				\hline
				\hline
				\multicolumn{1}{|c|}{} &\multicolumn{3}{|c|}{Rayleigh} & \multicolumn{2}{|c|}{Rician} \\  \hline
				\multicolumn{1}{|c|}{Number of tags} &\multicolumn{1}{|c|}{GMM} & \multicolumn{1}{|c|}{FNN} & \multicolumn{1}{|c|}{CNN}  & \multicolumn{1}{|c|}{FNN} & \multicolumn{1}{|c|}{CNN} \\  \hline
				\multicolumn{1}{|c|}{1 tag}& \multicolumn{1}{|c|}{0.9988} &\multicolumn{1}{|c|}{1.0} & \multicolumn{1}{|c|}{1.0}  &\multicolumn{1}{|c|}{1.0} & \multicolumn{1}{|c|}{1.0} \\ \hline
				\multicolumn{1}{|c|}{2 tags}& \multicolumn{1}{|c|}{0.9855} &\multicolumn{1}{|c|}{0.993} & \multicolumn{1}{|c|}{1.0}  &\multicolumn{1}{|c|}{0.988} & \multicolumn{1}{|c|}{1.0} \\ \hline
				\multicolumn{1}{|c|}{3 tags}& \multicolumn{1}{|c|}{0.9688} &\multicolumn{1}{|c|}{0.889} & \multicolumn{1}{|c|}{0.992}  &\multicolumn{1}{|c|}{0.813} & \multicolumn{1}{|c|}{0.991} \\ \hline
				\multicolumn{1}{|c|}{4 tags}& \multicolumn{1}{|c|}{0.863} &\multicolumn{1}{|c|}{0.863} & \multicolumn{1}{|c|}{0.894} &\multicolumn{1}{|c|}{0.831} & \multicolumn{1}{|c|}{0.89} \\ \hline			
				\multicolumn{1}{|c|}{Overall}& \multicolumn{1}{|c|}{0.954} &\multicolumn{1}{|c|}{0.936} & \multicolumn{1}{|c|}{0.971} &\multicolumn{1}{|c|}{0.908} & \multicolumn{1}{|c|}{0.97} \\ \hline 								
				\hline	
		\end{tabular}}
		\label{tab2} 	
		\vspace{-0.25cm}	
	\end{table}
	
	The MMSE loss values of channel parameters are presented in Table \ref{tab3} for tag numbers up to $4$ at an SNR value of 20 dB under both channel models. It is demonstrated that the loss values can be down to $10^{-4}$ and the network can extract the channel coefficients successfully with the help of preamble symbols. The same observation between two channel models can be made for this case as well which shows the generalization capability of the network. 
	
	We also note that the neural networks for number of tags and channel estimation is trained off-line using extensive number of simulated samples, however, the practice of the trained models only require simple calculations (matrix multiplication and non-linear activation). Hence, the use of deep learning tools is feasible for real time implementation while this is not the case for GMM. GMM needs to calculate well performing parameters for each sample individually which makes it demanding for computational power and disadvantageous in terms of applicability in real time. 
				
	\begin{table}[h]
		\small
		\caption{Comparison of MMSE values for channel parameters at SNR value of 20 dB under both channel models.}
		\centering
		\renewcommand{\arraystretch}{1.2} 
		\begin{tabular}{|c|c|c|}
			\hline
			\hline					
			\multicolumn{1}{|c|}{Number of tags}  & \multicolumn{1}{|c|}{Rayleigh}  & \multicolumn{1}{|c|}{Rician} \\  \hline
			\multicolumn{1}{|c|}{1 tag} &\multicolumn{1}{|c|}{$6.71 \times 10^{-5}$} & \multicolumn{1}{|c|}{$6.96 \times 10^{-5}$}  \\ \hline
			\multicolumn{1}{|c|}{2 tags}&\multicolumn{1}{|c|}{$1.09 \times 10^{-4}$} & \multicolumn{1}{|c|}{$1.09 \times 10^{-4}$} \\ \hline
			\multicolumn{1}{|c|}{3 tags}&\multicolumn{1}{|c|}{$1.45 \times 10^{-4}$} & \multicolumn{1}{|c|}{$1.51 \times 10^{-4}$} \\ \hline
			\multicolumn{1}{|c|}{4 tags}&\multicolumn{1}{|c|}{$1.65 \times 10^{-4}$} & \multicolumn{1}{|c|}{$1.68 \times 10^{-4}$} \\ \hline											
			\hline	
		\end{tabular}
		\label{tab3} 	
		\vspace{-0.4cm}	
	\end{table}

	\subsection{Throughput Performance}
	We now depict the throughput performance of the proposed collision recovery algorithm. We first consider the scenario where the receiver can only decode one tag from the collision ($J=1$), and then, extend our examples to the scenario where the receiver is capable of decoding up to $4$ tags ($J = 4$). For both scenarios, we assume that the receiver is equipped with single antenna and the channel is modeled as Rayleigh fading. We also perform simulations for Rician fading, however the results are quite similar to the Rayleigh fading since the trained models perform well for this case as well, hence, they are omitted. We perform our simulations over different SNR values under the optimal frame to tag ratio ($K/N$). 
	
\begin{figure}[b]
		\centering
		\includegraphics[clip, trim=0.9cm 0.05cm 1.8cm 1.1cm,width=1\columnwidth]{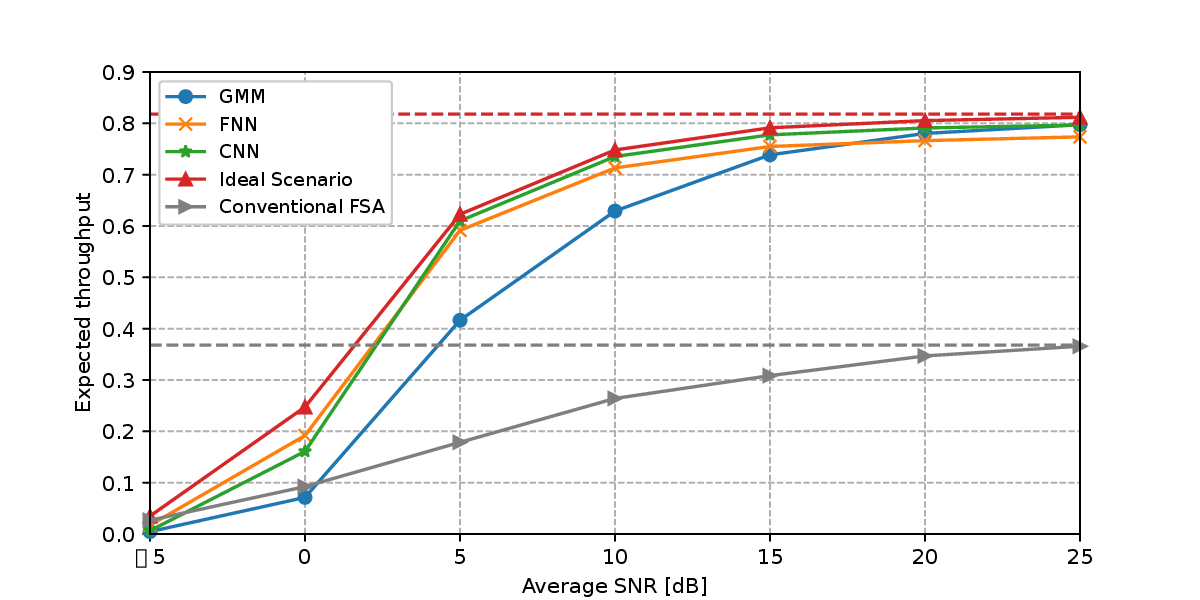}
		\caption{Throughput performance comparison of GMM, FNN and CNN along with the ideal scenario ($M=4,J=1$) and conventional FSA. }
		\label{fig9}
		\vspace{-0.6cm}
	\end{figure}

	In the first simulation setup, the receiver can only decode one tag from the collision up to $4$ tags ($M=4,J=1$). Fig. \ref{fig9} illustrates throughput performances of GMM, FNN and CNN over various SNR values. For comparison purposes, we also include the throughput values of the ideal scenario where the number of tags and channel coefficients are known perfectly by the receiver and conventional FSA algorithm. The dashed lines indicate the theoretical throughput of conventional FSA and FSA with collision recovery. We observe that in low SNR regime, GMM performance suffers significantly, both FNN and CNN perform well. However, this gap vanishes as SNR increase and all models provide similar performance to each other. 
	
\begin{figure}
		\centering
		\includegraphics[clip, trim=0.9cm 0.05cm 1.8cm 1.1cm,width=1\columnwidth]{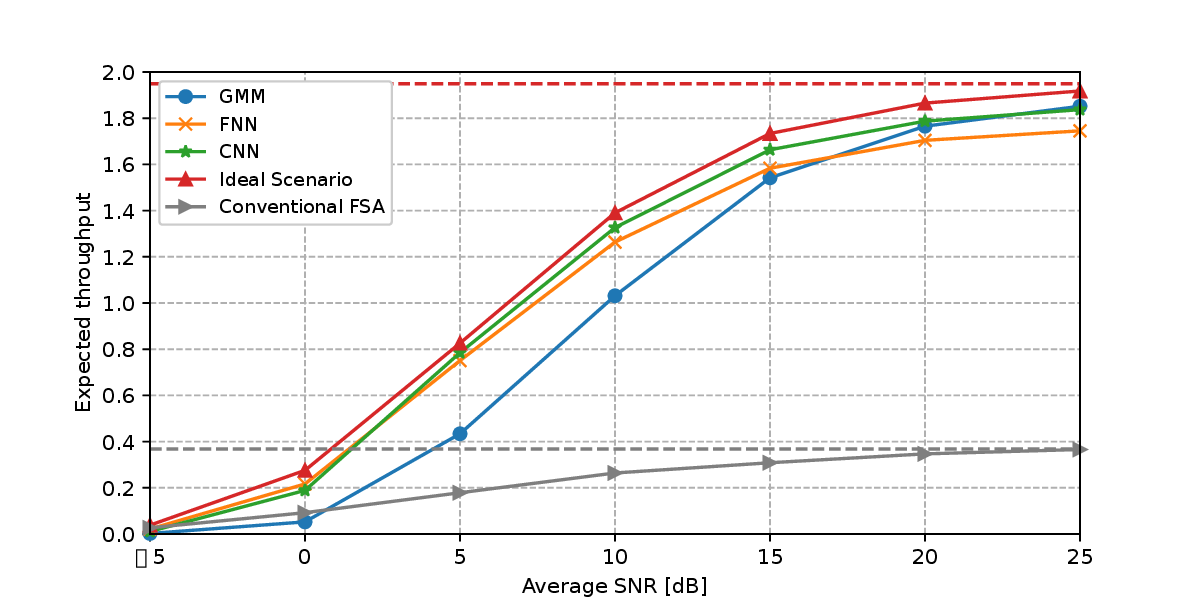}
		\caption{Throughput performance comparison of GMM, FNN and CNN along with the ideal scenario ($M=4,J=4$) and conventional FSA. }
		\label{fig10}
		\vspace{-0.6cm}
	\end{figure}

	In the second simulation setup, we consider the case where the receiver decode up to $4$ tags out of $4$ colliding tags ($M=4,J=4$). We depict the throughput values of GMM, FNN and CNN in Fig. \ref{fig10} along with the ideal scenario and conventional FSA. The throughput gain compared to the $J=1$ scenario is easy to observe and throughput value nearly approaching to $2$. The observations are similar to the previous setup and CNN still outperforming the other alternatives. 
	
	Finally, we compare our results with the existing works in the literature with single antenna receiver and present the maximum throughput values of each work in Table \ref{tab4}. The work \cite{angerer2010rfid} considers only $2$ colliding tags and recovery of a single tag ($M=2,J=1$) with zero-forcing receiver, thus, the throughput performance is lower than the other works. The authors in \cite{kaitovic2012channel} employ an MMSE receiver which is capable of recovering $2$ colliding tags ($M=2,J=2$) and obtain a throughput value of $0.841$. In \cite{tan2016collision}, voltage clustering is applied for the constellation points with a throughput value of $0.85$ under the simulation setup $M=3,J=2$. With our proposed approach, we show that it is possible to decode $4$ tags out of $4$ colliding ones ($M=4,J=4$) by improving throughput value to the $1.837$. 
	
	\begin{table}[h]
		\scriptsize
		\caption{Comparison of our work with the existing ones.} 
		\centering 
		\renewcommand{\arraystretch}{1.2} 
			\begin{tabular}{|c | c | c | c| }  
				\hline 
				\hline
				Work  		& Maximum Throughput & Method \& Setup	 	 				\\ [0.5ex]\hline
				\cite{angerer2010rfid}  & 0.587		& Zero-Forcing Receiver ($M=2,J=1$)	\\ [0.5ex]\hline 
				\cite{kaitovic2012channel}  & 0.841 	& MMSE Receiver	 	($M=2,J=2$) 	\\ [0.5ex]\hline 
				\cite{tan2016collision}  	& 0.85  	& Voltage Clustering ($M=3,J=2$) 		\\ [0.5ex]\hline 
				\textbf{Our work}  					& \bftab 0.797 	& \textbf{ML-based recovery} ($M=4,J=1$)  \\ [0.5ex] \hline 	
				\textbf{Our work}  					& \bftab 1.837 	& \textbf{ML-based recovery} ($M=4,J=4$)  \\ [0.5ex] \hline 			
				\hline
			\end{tabular}
			\label{tab4}
			\vspace{-0.4cm}	
		\end{table}
		
		\section{Conclusions and Future Work}
		We have considered the problem of recovering tags in collision for UHF-RFID systems. Different from the existing recovery methods, we propose a machine learning based algorithm. We show that the proposed models can estimate the number of tags and the channel coefficients successfully along with a suitable design and training. The trained models enable the receiver to recover tag signals from the collided one by using minimum distance decoding. We also perform simulations to demonstrate the performance of the proposed approach which provides a significant improvement in throughput.
		
		As a future work, one direction might be to confirm the validity of the proposed method on measurement data obtained from the real time implementations. In our work, we consider a simplified communication model with coherent detection to show the applicability of deep learning tools for UHF-RFID systems and it is possible to extend our approach for more realistic setups taking the tag asynchronism and different symbols periods into account. In addition, we study a system model where the receiver is equipped with only a single antenna which can be generalized to the multiple antenna setup. Another line of work might be to examine the complexity of the trained models and how they can be coupled with an experimental configuration. 	
		
		\bibliographystyle{IEEEtran}
		\bibliography{ref} 
		
	\end{document}